\documentclass[osajnl,twocolumn,showpacs,superscriptaddress,10pt]{revtex4-1}
\usepackage{xcolor}
\usepackage{citesort}
\usepackage{graphicx}
\usepackage{dcolumn}
\usepackage{bm}
\usepackage{amssymb}
\usepackage{amsmath}
\usepackage{mathrsfs}
\usepackage{subfigure}
\usepackage{setspace}
\usepackage[mathlines]{lineno}
\begin{document}

\title{Fringe-locking method for the weak equivalence principle test by simultaneous dual-species atom interferometers}

\author{Xiao-Chun Duan}
\author{Xiao-Bing Deng}
\author{De-Kai Mao}
\author{Min-Kang Zhou}
\author{Cheng-Gang Shao}
\author{Zhong-Kun Hu}
\email{zkhu@hust.edu.cn}
\affiliation{MOE Key Laboratory of Fundamental Physical Quantities
Measurements, School of Physics, Huazhong University of Science and
Technology, Wuhan 430074, People's Republic of China}

\begin{abstract}
We theoretically investigate the application of the fringe-locking
method (FLM) in the dual-species quantum test of the weak
equivalence principle (WEP). With the FLM, the measurement is
performed invariably at the midfringe, and the extraction of the
phase shift for atom interferometers is linearized. For the
simultaneous interferometers, this linearization enables a good
common-mode rejection of vibration noise, which is usually the main
limit for high precision WEP tests of dual-species kind. We note
that this method also allows for an unbiased determination of the
gravity accelerations difference, which meanwhile is readily to be
implemented.
\end{abstract}


\maketitle

\section{Introduction}
\label{intro}

Benefit from the highly developed atom interferometry technology,
cold atoms, possing both internal and external degrees of freedom, become
ideal probes in many precision measurements. They have been
successfully used in measuring gravity acceleration \cite{Kas92,Pet99,Pet01,Le08,Mul08,Zhou12,Bid13,Alt13,Hu13,Gil14},
gravity gradient \cite{Bid13,Alt13,Sna98,McG02,Sor14,Yu06,Sor12,Duan14}, rotation \cite{Len97,Gus97,Wu07,Gau09,Sto11,Tac12,Rak14},
magnetic field gradient \cite{Davis08,Zhou10,Hu11,Bar11}, etc. Atom interferometers also play an important
role in fundamental physics, such as the measurement of fine
structure constant, the determination of gravitational constant $G$
\cite{Fix07,Lam08,Rosi14}, and the test of the weak equivalence
principle (WEP) \cite{Pet99,Fray04,Bon13,Bon15,Sch14,Tar14,Zhou15}.

WEP, as one of the cornerstones of Einstein's general relativity,
states that all masses fall in the same way in a gravitational field
regardless of their internal structure and composition.
Verifications of WEP using macroscopic masses have achieved a level
of 10$^{-13}$ \cite{Wil04,Sch08}, while the best level for testing
WEP on quantum basis is at the level of 10$^{-9}$ \cite{Pet99}.
Testing WEP using microscopic particles still stimulated wide
interest since the neutron interferometer \cite{Col75}. The
reason is that, quantum objects offer more possibilities to break
WEP and meanwhile they also afford potentially higher precision and
well defined properties \cite{Sch14,Lam98,Dim08}. Up to date, WEP
tests on quantum basis have been performed using atoms versus
macroscopic bodies \cite{Pet99} or other atom specie \cite{Sch14}, using single-specie atoms in
different hyperfine levels \cite{Fray04} or in different spin
orientations \cite{Duan15}, and using bosonic atoms versus fermionic
atoms \cite{Tar14}, achieving a level of 10$^{-7}$. In these tests,
the corresponding gravity accelerations are usually independently
measured and then compared, in which situation the vibration noise
cannot be common-mode rejected. Under this circumstance, WEP test
using simultaneous dual-species atom interferometers is particular
interesting for its intrinsic capability of common-mode suppressing
the vibration noise \cite{Bon13,Zhou15}. WEP tests of this kind have
already been performed by several groups, achieving a level of
10$^{-8}$ \cite{Zhou15}, and tests with higher precision are under
development \cite{Sug13,Kuhn14} or have been proposed \cite{Agu10,Alt15}.

However, common-mode rejection of vibration noise in WEP tests of
dual-species kind, especially using non-isotope species, is not so
direct as that, for example, in atom gradiometers. In the operation
of coupled interferometers using the same atoms, it is mature to
extract the differential phase shift $\Delta \Phi$ in a common-mode
noises immune way by the ellipse fit method \cite{Fos02} or Bayesian
estimation \cite{Sto07,Var09,Chen14}. But in WEP tests of
dual-species kind, the scale factors ${S_j} \cong
k_j^{{\rm{eff}}}T_j^2$ are usually different for the two
interferometers, which originates from the different effective Raman
wave vectors used for the atom species $j=$ 1 and 2 when the pulses
separation times $T_j$ are the same. The different scale factors
cause two aspects of complexity in the signal extraction of WEP
tests. Firstly, the interested signal $\Delta g \equiv {g_1} -
{g_2}$ is not proportional to the differential phase shift, which
excludes the direct extraction of $\Delta g$ using usual ellipse fit
method or Bayesian estimation. Secondly, the induced phase
fluctuations by vibration noise are not the same for the two
interferometers, which increases the difficulty to common-mode
suppress the vibration noise. The scale factors can be made the same
by using different values for the pulses separation times $T_j$,
which is particularly favorable in the case of the ratio $r \equiv
k_2^{{\rm{eff}}}/k_{\rm{1}}^{{\rm{eff}}}$ close to unity
\cite{Var09}. However, in this situation, the experienced vibration
noise is not exactly the same for the two interferometers, which
thus excludes the possibility of perfect common-mode suppression. It
is also proposed to simultaneously measure the vibration noise  by
an auxiliary sensor and then reconstruct the fringes \cite{Per15,Bar15}.
The effectiveness of this method depends on the quality of
the correlation between real and measured vibrations, which is hard
to ensure when the aimed precision of the WEP test is beyond the
intrinsic noise of state-of-art vibrations sensors. This problem may
also be mathematically resolved by improved ellipse fit method,
Bayesian estimation or direct phase extraction \cite{Chen14,Bon15,Bar15}.
However, these solutions either require complex computation
(sometimes even causing a bias result) or suffer from the trouble of
separating the WEP violation signal from total differential phase
shift.

It is already clear that the gravity acceleration or the vibration
induced phase shift is linear to $k_j^{{\rm{eff}}}$, and we further
note it is the conventional non-linear phase extraction process from
the interference fringe that complicates the WEP violation signal
separation and the vibration noise common-mode rejection. Actually,
the fringe-locking method (FLM) has already been adopted formerly
for single atom interferometers \cite{Cla95,Che06,Mer09,Zhou13}
and recently for coupled interferometers \cite{Duan14}, by
which the signal extraction of the interferometer can be linearized
as already exploited in single atom interferometer. In this work, we
propose to operate the dual-species atom interferometers in the
fringe-locking mode to linearize the signal extraction, which
promises a good common-mode rejection of vibration noise, especially
in the case of low level vibration noise. We note that, in the WEP
test of dual-species kind, it only needs a change in the control of
the Raman lasers effective frequencies to perform the fringe-locking
method, and the corresponding signal extraction is quite direct.
Moreover, this fringe-locking method allows for an unbiased
determination of $\Delta g$.

\section{Review of Fringe-Locking Method}
\label{Fringelocking}

\begin{figure}[tbp]
\centering
\includegraphics[trim=40 10 40 20,width=0.40\textwidth]{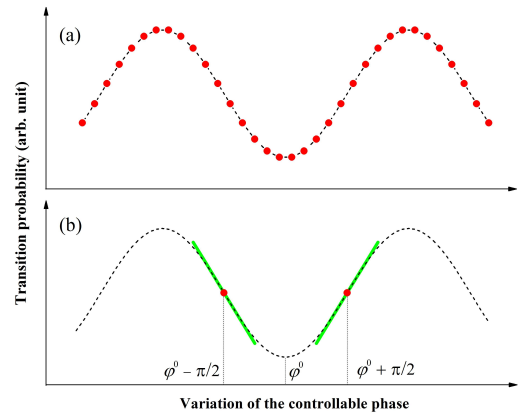}
\caption{\label{fig:1}(color online) Different methods to operate
the interferometer, with (a) for recording full fringes and (b) for
performing measurements at midfringe. In the conventional method, the phase is
scanned step by step, while in the FLM, the phase is modulated by
$\pm \pi /2$ with respect to an appropriate center, denoted as
$\varphi ^0$ here. The dash lines are here to guide the eyes. And
the two thick green lines are the tangent lines of the midfringe,
which show good approximation for the fringe near the center.}
\end{figure}

In light pulses atom interferometers, the interference pattern is
usually manifested as the variation of the transition probability
between the two ground levels of the atom. And the transition
probability $P$ is typically expressed as \cite{Kas92,Pet01}
\begin{equation}
P = A + B\cos (\varphi  + {\varphi _{\rm{m}}}), \label{Eq1}
\end{equation}
which forms a cosine fringe when the controllable $\varphi$ is
scanned. In Eq. (\ref{Eq1}),  $A$ is the fringe offset, $B$ is the fringe
amplitude, and $\varphi _{\rm{m}}$ indicates the phase shift induced
by the physical quantity to be measured. In conventional method, a
full cosine fringe is obtained by scanning $\varphi$ step by step,
as shown in Fig. \ref{fig:1}(a), and ${\varphi _{\rm{m}}}$ is then acquired by
a cosine fitting. Alternatively, the FLM can be adopted
\cite{Cla95,Che06,Mer09}, as shown in Fig. \ref{fig:1}(b). In the
fringe-locking approach, the appropriate value of $\varphi$, denoted
as ${\varphi ^0}$, is found to make $\varphi + {\varphi
_{\rm{m}}}\sim 2n\pi$ (where $n$ is an integer). Then $\varphi$ is
modulated by $\pm \pi /2$ with respect to the center of ${\varphi
^0}$ so that the measurement is always performed at midfringe,
alternately to the right and to the left side of the central fringe.
In this case, the transition probabilities $P$ for every two
consecutive launches can be expressed as
\begin{subequations}
\label{Eq2}
\begin{equation}
{P[2l-1]} = A + B\cos ({\varphi ^0} + {\varphi _{\rm{m}}} - \pi /2),\label{subeq2:1}
\end{equation}
\begin{equation}
{P[2l]} = A + B\cos ({\varphi ^0} + {\varphi _{\rm{m}}} + \pi /2),\label{subeq2:2}
\end{equation}
\end{subequations}
where $l$ ($l = 1,2,3, \cdots$)denotes the index of modulating
cycles. The difference between $P[2l-1]$ and $P[2l]$ could be used
as a criterion of whether $\varphi^0 + {\varphi _{\rm{m}}}= 2n\pi$
is fulfilled, which thus enables feedback control of $\varphi^0$.
According to the linear approximating of Eq. (\ref{Eq2}) at midfringe, the
correction can be expressed as
\begin{equation}
\delta ({\varphi ^0}) = ({P[2l]} - {P[2l-1]})/2B, \label{Eq3}
\end{equation}
where the fringe amplitude $B$ must be known in advance by scanning
full fringe. Once the corrections are made to form a closed feedback
loop, the equation ${\varphi ^0} + {\varphi _{\rm{m}}} = 2n\pi$ is
supposed to be enable, from which the value of the interested
${\varphi _{\rm{m}}}$ can be deduced from ${\varphi ^0}$ in a linear
way. It is shown in Eq. (\ref{Eq3}) that when the interferometers is
operated at the midfringe, the measurement works in a linear region.
This linearization character is helpful for signal extraction in the
WEP test of dual-species kind, as will be discussed detailedly in the following
section.

\section{Application in WEP Test}
In the above section, the fringe-locking approach for single
interferometer is illuminated, and the Raman laser phase is
convenient to control, which thus usually plays the role of
${\varphi}$. Generally speaking, one needs two independent
controllable phases to simultaneously lock two interferometers. In
Ref. 16, in addition to Raman laser phase, the phase
shift due to magnetic field gradient is explored as the other
controllable phase for an atom gravity gradiometer. In WEP test
using simultaneous dual-species atom interferometers, there are
already two independent groups of Raman lasers (usually one group
used for one specie atom), and thus it is natural to explore the two
controllable Raman laser phases for fringe locking.

For each specie atom interferometer using $\pi /2 - \pi  - \pi /2$
Raman pulses scheme, the controllable phase ${\varphi}$, namely the
Raman lasers phase, can be expressed as \cite{Chen14,Bon15,Bar15,Li15}
\begin{equation}
{\varphi _j} = {\alpha _j}({T_j} + 2{\tau _j})({T_j} + 4{\tau
_j}/\pi ) \equiv {\alpha _j}{f_{{\rm{DC}}}}({T_j},{\tau _j}),
\label{Eq4}
\end{equation}
where $\alpha _j$ is the chirp rate of the effective Raman laser
frequency used to compensate the Doppler shift due to gravity, and
${\tau _j}$ is the $\pi/2$ pulse duration. With the Raman pulses
duration effect neglected, ${f_{{\rm{DC}}}}({T_j},{\tau _j})$ just
simplifies to $T_j^2$. And the interested phase ${\varphi
_{\rm{m}}}$, namely the phase related to the gravity acceleration,
can be expressed as \cite{Chen14,Bon15,Bar15,Li15}
\begin{equation}
{({\varphi _{\rm{m}}})_j} =
k_j^{{\rm{eff}}}{g_j}{f_{{\rm{DC}}}}({T_j},{\tau _j}), \label{Eq5}
\end{equation}
where the gravity acceleration of $j$ specie atoms is denoted as
$g_j$ to account for possible WEP violation. In order to clearly
manifest the ability of common-mode rejection with FLM, the phase
due to vibration is explicitly included in the total phase shift,
which can be expressed as \cite{Chen14,Bon15,Bar15,Li15}
\begin{eqnarray}
{({\varphi _{{\rm{vib}}}})_j}&& = k_j^{{\rm{eff}}}\int_{ - \infty
}^\infty  {\xi(t - {{({t_i})}_j})\int_{ - \infty }^t {a(t')dt'} } dt \nonumber\\
 &&\equiv k_j^{{\rm{eff}}}h({T_j},{\tau _j};a(t),{({t_i})_j}),
\label{Eq6}
\end{eqnarray}
where $\xi(t)$ is the sensitivity function \cite{Che08}, and $t_i$
is the central time of the interfering progress for the $i$-th shot
measurement. We note that both ${f_{{\rm{DC}}}}({T_j},{\tau _j})$
and $h({T_j},{\tau _j};a(t),{t_i})$ are irrelative to
$k_j^{{\rm{eff}}}$ \cite{Chen14,Bon15,Bar15,Li15}. And in the case of
identical pulses separation time (namely ${T_1} = {T_2} \equiv T$),
identical effective Rabi frequencies (thus ${\tau _1} = {\tau _2}
\equiv \tau$), and simultaneous interferometers (thus ${({t_i})_1} =
{({t_i})_2} \equiv {t_i}$ as well as experiencing the same vibration
noise $a(t)$), they are identical for the dual-species atom
interferometers. This is exactly the situation we hope (and also are
able to) manage to achieve, and here we then abbreviate
${f_{{\rm{DC}}}}(T,\tau )$ and $h(T,\tau ;a(t),{t_i})$ as
$f_{\rm{DC}}$ and ${h[i]}$, respectively.

Once the appropriate value of Raman laser phase, denoted as $\varphi
_j^0$, are found to make ${\varphi _j} + {({\varphi
_{\rm{m}}})_j}\sim 2n_j\pi$ for each interferometers, the Raman
laser phases are then modulated by $\pm \pi /2$. The corresponding
transition probabilities for the two interferometers for every two
consecutive launches can be expressed as
\begin{subequations}
\label{Eq7}
\begin{eqnarray}
{P_j[2l-1]} = &&{A_j} + {B_j}\cos (\varphi _j^0 + k_j^{{\rm{eff}}}{g_j}{f_{{\rm{DC}}}}\nonumber\\
&&+ k_j^{{\rm{eff}}}{h[2l-1]} - \pi /2),\label{subeq7:1}
\end{eqnarray}
\begin{eqnarray}
{P_j[2l]} =&&{A_j} + {B_j}\cos (\varphi _j^0+k_j^{{\rm{eff}}}{g_j}{f_{{\rm{DC}}}}\nonumber\\
&&+ k_j^{{\rm{eff}}}{h[2l]}+ \pi /2).\label{subeq7:2}
\end{eqnarray}
\end{subequations}
And as same as in single interferometer, the correction is made as
\begin{equation}
\delta ({(\varphi _j^0)[l]}) = ({P_j [2l]} - {P_j[2l-1]})/2{B_j},
\label{Eq8}
\end{equation}
where $l$ denotes the $l$-th correction (the index of $\varphi _j^0$
is explicitly indicated here). The $(l+1)$-th phase modulation
center will be ${(\varphi _j^0)[l +1]} ={(\varphi _j^0)[l]} + \delta
({(\varphi _j^0)[l]})$. And for single interferometer, the $l$-th
measured value of the gravity acceleration is then expressed as
${({g_{{\rm{meas}}}})_j}[l] = (2{n_j}\pi  - \varphi _j^0[l]-\delta
({(\varphi _j^0)[l]}))/k_j^{{\rm{eff}}}{f_{{\rm{DC}}}}$. This
measured value is obviously affected by the vibration noise, which
can be actually explicitly deduced from the linear approximation of
Eq. (\ref{Eq7}) at midfringe, namely
\begin{equation}
{({g_{{\rm{meas}}}})_j}[l] = {g_j} + (h[2l-1] +
h[2l])/2{f_{{\rm{DC}}}}. \label{Eq9}
\end{equation}
However, the measured WEP violation signal is the difference of the
measured gravity accelerations, which is then expressed as
\begin{equation}
{(\Delta g)_{{\rm{meas}}}}[l] \equiv {({g_{{\rm{meas}}}})_1}[l] -
{({g_{{\rm{meas}}}})_2}[l] = {g_1} - {g_2}, \label{Eq10}
\end{equation}
which is exactly the possible WEP violation signal one searches for.
It is clearly shown from Eq. (\ref{Eq10}) that the FLM promises a perfect
common-mode rejection of vibration noise within the first order
approximation for WEP tests using simultaneous dual-species atom
interferometers. And this common-mode rejection capability profits
from the linearized signal extraction.

\section{Simulation and Result}
\label{Simulation}

Eq. (\ref{Eq9}) is based on the linear approximation of the measurement
equation Eq. (\ref{Eq7}), and so is the consequent common-mode rejection.
Actually, vibration noise would cause a departure of the measurement
point from midfringe, which would then affect the linear
approximation. It is easily imaged that this affection will increase
with the noise level. And since the vibration noise induced phases
for the two interferometers are different, the relative sites of the
measurement points at respective fringe are also different, which
will limit the common-mode rejection capability of this
fringe-locking method. The rejection capability will be investigated
here by numerical simulation. For simplicity, in the simulation the
transition probabilities are re-written as
\begin{subequations}
\label{Eq11}
\begin{equation}
{P_1} = {A_1} + {B_1}\cos ({\varphi _1} + {({\varphi _g})_1} + {\varphi _{{\rm{vib}}}}),\label{subeq11:1}
\end{equation}
\begin{equation}
{P_2} = {A_2} + {B_2}\cos ({\varphi _2} + r{({\varphi _g})_2} + r{\varphi _{{\rm{vib}}}}),\label{subeq11:2}
\end{equation}
\end{subequations}
where ${({\varphi _g})_1}$ is equivalent to ${({\varphi
_{\rm{m}}})_1}$, and ${({\varphi _g})_2}$ is defined as ${({\varphi
_g})_2} \equiv k_1^{{\rm{eff}}}{g_2}{f_{{\rm{DC}}}}$. With this
redefinition, ${({\varphi _g})_1}={({\varphi _g})_2}$ in the absence
of WEP violation. The vibration noise is simulated by randomly
drawing the values of ${\varphi _{{\rm{vib}}}}$ in a Gaussian
distribution with a standard deviation of ${\sigma _{{\varphi
_{{\rm{vib}}}}}}$. To simulate the fringe locking, the Raman laser
phase ${\varphi _1}$ (${\varphi _2}$) is modulated by $\pm \pi /2$
with a center of $\varphi _1^0$ ($\varphi _2^0$), and the
corrections for every two consecutive launches is made as Eq. (\ref{Eq8})
(The initial values of ${\varphi _1}$ and ${\varphi _2}$ are $-
{({\varphi _g})_1}$ and $- r{({\varphi _g})_2}$, respectively). The
corresponding measured phase shifts due to gravity acceleration are
then $({\varphi _g})_1^{{\rm{meas}}} = - \varphi _1^0$ and
$({\varphi _g})_2^{{\rm{meas}}} =  - \varphi _2^0/r$, respectively,
from which the interested possible WEP violation signal can be
deduced as $\Delta {\Phi _g} \equiv{({\varphi _g})_1} - {({\varphi
_g})_2} =  - (\varphi _1^0 - \varphi _2^0/r)$.

In the numerical simulation, the dependence of the common-mode
rejection efficiency on the vibration noise level (characterized by
${\sigma _{{\varphi _{{\rm{vib}}}}}}$) and the ratio of the
effective Raman wave vector (characterized by $r$) is investigated.
For each ${\sigma _{{\varphi _{{\rm{vib}}}}}}$ and $r$, 10$^4$
modulation cycles are simulated for the two interferometers, in
which process $2\times 10^4$ pairs $(P_1, P_2)$ are generated with
$A_1=A_2=0.5$ and $B_1=B_2=0.5$ and 10$^4$ pairs ($({\varphi
_g})_1^{{\rm{meas}}}, ({\varphi _g})_2^{{\rm{meas}}}$) are obtained.
The influence of the vibration noise on the interested signal is
characterized by the Allan deviation of the $\Delta {\Phi _g}$,
which is calculated by the obtained pairs ($({\varphi
_g})_1^{{\rm{meas}}}, ({\varphi _g})_2^{{\rm{meas}}}$). Since the
simulated vibration noise is white noise, the calculated Allan
deviation of $\Delta {\Phi _g}$, denoted as ${\sigma _{\Delta {\Phi
_g}}}(N)$, scales down by the inverse square-root of the number of
measurement, namely ${\sigma _{\Delta {\Phi _g}}}(N) = \sigma
_{\Delta {\Phi _g}}^1/\sqrt N$. Here $N$ denotes the number of
measurements, and $\sigma _{\Delta {\Phi _g}}^1$ stands for the
measurement sensitivity, which is obtained by white noise model
fitting of ${\sigma _{\Delta {\Phi _g}}}(N)$ versus $N$.

\begin{figure}[tbp]
\centering
\includegraphics[trim=40 10 40 20,width=0.40\textwidth]{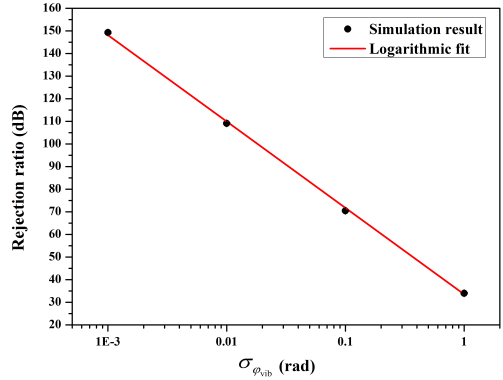}
\caption{\label{fig:2}(color online) Variation of the rejection
ratio along the vibration noise level. In this group of simulation,
the gravity induced phase is set as ${({\varphi _g})_1} = {({\varphi
_g})_2} = 2$ rad, and the ratio for the effective Raman wave vector
is set as 780/767, the value for the dual-species of $^{39}$K versus
$^{87}$Rb.}
\end{figure}

The efficiency of the common-mode noise rejection is characterized
by the rejection ratio  ${\sigma _{{\varphi _{{\rm{vib}}}}}}/\sigma
_{\Delta {\Phi _g}}^1$. The simulation result for the dependence of
the rejection ratio on the vibration noise level ${\sigma _{{\varphi
_{{\rm{vib}}}}}}$ is shown in Fig. \ref{fig:2}. According to the logarithmic
fit in Fig. \ref{fig:2}, the rejection ratio scales by $-38.2(1)$ dB per
octave, which is very close to the expected value of $-40$ dB per
octave as a result of the third order expansion of the sinusoid
function. This means that the common-mode noise rejection capability
of the fringe-locking method improves quite fast as the vibration
noise decreases.  This simulation also tells that even if the
vibration noise is as large as 1 rad, there is still about 33 dB
rejection ratio for the $^{39}$K versus $^{87}$Rb dual-species
interferometers.

\begin{figure}[tbp]
\centering
\includegraphics[trim=40 10 40 20,width=0.40\textwidth]{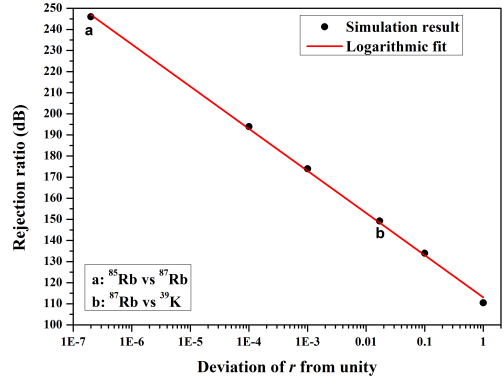}
\caption{\label{fig:3}(color online) Dependence of the rejection
ratio on the ratio of the effective Raman wave vector. The ratios
for dual-species of $^{87}$Rb versus $^{85}$Rb and $^{39}$K versus
$^{87}$Rb are explicitly displayed. In this group of simulation, the
gravity induced phase is set as ${({\varphi _g})_1} = {({\varphi
_g})_2} = 2$ rad, and the vibration noise level is fixed at ${\sigma
_{{\varphi _{{\rm{vib}}}}}} = 1$ mrad.}
\end{figure}

The simulation for different $r$ is then performed with a fixed
vibration noise level of ${\sigma _{{\varphi _{{\rm{vib}}}}}} = 1$
mrad. The result is shown in Fig. \ref{fig:3}, which also shows a
logarithmic relation. According to the logarithmic fit, the
rejection ratio scales down by $-19.98(4)$ dB per octave. It is
shown that even if the deviation of the ratio from unity is as large
as 1, there is still about 110 dB rejection ratio for ${\sigma
_{{\varphi _{{\rm{vib}}}}}} = 1$ mrad.

In addition to common-mode noise rejection capability, the bias of
the extracted $\Delta {\Phi _g}$ is also much concerned. We have
checked that the average value of $\Delta {\Phi _g}$ extracted by
this fringe-locking method is exactly equal to the differential
value of the set ${({\varphi _g})_1} $ and ${({\varphi _g})_2} $,
whatever the vibration noise level (as long as ${\sigma _{{\varphi
_{{\rm{vib}}}}}} \leq 1$ rad) and the ratio of the effective Raman
wave vector are.

\section{Discussion and Conclusion}
\label{discuss}

The simulation for different fringe amplitudes of the two
interferometers (namely ${B_1} \ne {B_2}$) is also performed, and
the dependence of the rejection ratio on ${\sigma _{{\varphi
_{{\rm{vib}}}}}}$ and $r$ doesn't change. This is consistent with
our knowledge. In the presence of only phase noises, the absolute
value of the amplitude doesn't matter much for the FLM as long as it
is exactly known.  In actual experiment, the fringe amplitudes of
the two interferometers are pre-determined by scanning the full
fringe and then performing cosine fittings. In the fringe-locking
mode, the realtime information about the fringe amplitudes is lost,
of which possible drift will affect the fringe-locking. This can be
resolved by occasionally switching back to the full-fringe recording
mode to get renewed fringe amplitudes.

In conclusion, we have shown the capability of common-mode noise
rejection using fringe-locking method in WEP test by simultaneous
dual-species atom interferometers. We note that it is quite
convenient to perform this method in the dual-species
interferometers, and it is also quite direct to extract the signal.
Of importance, this signal extraction approach allows for an
unbiased determination of the gravity accelerations difference, and
affords a good common-mode noise rejection, especially in the low
vibration noise level. This will thus greatly alleviate the demand
for the vibration noise isolation in WEP test of dual-species kind.

\section*{Acknowledgements}

This work is supported by the National Natural Science Foundation of
China (Grants No. 41127002, No. 11574099, and No. 11474115) and the
National Basic Research Program of China (Grant No. 2010CB832806).


\begin{thebibliography}{60}
\bibitem{Kas92} M. Kasevich and S. Chu, ¡°Atomic interferometry using stimulated raman transitions,¡± Phys. Rev. Lett. {\bf 67}, 181 (1991).
\bibitem{Pet99} A. Peters, K. Y. Chung, and S. Chu, ¡°Measurement of gravitational acceleration by dropping atoms,¡± Nature {\bf 400}, 849-852 (1999).
\bibitem{Pet01} A. Peters, K. Y. Chung, and S. Chu, ¡°High-precision gravity measurements using atom interferometry,¡± Metrologia {\bf 38}, 25 (2001).
\bibitem{Le08} J. Le Gou\"{e}t, T. Mehlst\"{a}ubler, J. Kim, S. Merlet, A. Clairon, A. Landragin, and F. P. Dos Santos, ¡°Limits to the sensitivity of a low noise compact atomic gravimeter,¡± Appl. Phys. B {\bf 92}, 133-144 (2008).
\bibitem{Mul08} H. M\"{u}ller, S.-w. Chiow, S. Herrmann, S. Chu, and K.-Y. Chung, ¡°Atominterferometry tests of the isotropy of post-newtonian gravity,¡± Phys. Rev. Lett. {\bf 100}, 031101 (2008).
\bibitem{Zhou12} M.-K. Zhou, Z.-K. Hu, X.-C. Duan, B.-L. Sun, L.-L. Chen, Q.-Z. Zhang, and J. Luo, ¡°Performance of a cold-atom gravimeter with an active vibration isolator,¡± Phys. Rev. A {\bf 86}, 043630 (2012).
\bibitem{Bid13} Y. Bidel, O. Carraz, R. Charri\`{e}re, M. Cadoret, N. Zahzam, and A. Bresson, ¡°Compact cold atom gravimeter for field applications,¡± Appl. Phys. Lett. {\bf 102}, 144107 (2013).
\bibitem{Alt13} P. Altin, M. Johnsson, V. Negnevitsky, G. Dennis, R. Anderson, J. Debs, S. Szigeti, K. Hardman, S. Bennetts, G. McDonald et al., ¡°Precision atomic gravimeter based on bragg diffraction,¡± New J. Phys. {\bf 15}, 023009 (2013).
\bibitem{Hu13} Z.-K. Hu, B.-L. Sun, X.-C. Duan, M.-K. Zhou, L.-L. Chen, S. Zhan, Q.-Z. Zhang, and J. Luo, ¡°Demonstration of an ultrahigh-sensitivity atominterferometry absolute gravimeter,¡± Phys. Rev. A {\bf 88}, 043610 (2013).
\bibitem{Gil14} P. Gillot, O. Francis, A. Landragin, F. P. Dos Santos, and S. Merlet, ¡°Stability comparison of two absolute gravimeters: optical versus atomic interferometers,¡± Metrologia {\bf 51}, L15 (2014).
\bibitem{Sna98} M. Snadden, J. McGuirk, P. Bouyer, K. Haritos, and M. Kasevich, ¡°Measurement of the earth¡¯s gravity gradient with an atom interferometerbased gravity gradiometer,¡± Phys. Rev. Lett. {\bf 81}, 971 (1998).
\bibitem{McG02} J. McGuirk, G. Foster, J. Fixler, M. Snadden, and M. Kasevich, ¡°Sensitive absolute-gravity gradiometry using atom interferometry,¡± Phys. Rev. A {\bf 65}, 033608 (2002).
\bibitem{Sor14} F. Sorrentino, Q. Bodart, L. Cacciapuoti, Y.-H. Lien, M. Prevedelli, G. Rosi, L. Salvi, and G. Tino, ¡°Sensitivity limits of a raman atom interferometer as a gravity gradiometer,¡± Phys. Rev. A {\bf 89}, 023607 (2014).
\bibitem{Yu06} N. Yu, J. Kohel, J. Kellogg, and L. Maleki, ¡°Development of an atominterferometer gravity gradiometer for gravity measurement from space,¡± Appl. Phys. B {\bf 84}, 647-652 (2006).
\bibitem{Sor12} F. Sorrentino, A. Bertoldi, Q. Bodart, L. Cacciapuoti, M. De Angelis, Y.-H. Lien, M. Prevedelli, G. Rosi, and G. Tino, ¡°Simultaneous measurement of gravity acceleration and gravity gradient with an atom interferometer,¡± Appl. Phys. Lett. {\bf 101}, 114106 (2012).
\bibitem{Duan14} X.-C. Duan, M.-K. Zhou, D.-K. Mao, H.-B. Yao, X.-B. Deng, J. Luo, and Z.-K. Hu, ¡°Operating an atom-interferometry-based gravity gradiometer by the dual-fringe-locking method,¡± Phys. Rev. A {\bf 90}, 023617 (2014).
\bibitem{Len97} A. Lenef, T. D. Hammond, E. T. Smith, M. S. Chapman, R. A. Rubenstein, and D. E. Pritchard, ¡°Rotation sensing with an atom interferometer,¡± Phys. Rev. Lett. {\bf 78}, 760 (1997).
\bibitem{Gus97} T. Gustavson, P. Bouyer, and M. Kasevich, ¡°Precision rotation measurements with an atom interferometer gyroscope,¡± Phys. Rev. Lett. {\bf 78}, 2046 (1997).
\bibitem{Wu07} S. Wu, E. Su, and M. Prentiss, ¡°Demonstration of an area-enclosing guided-atom interferometer for rotation sensing,¡± Phys. Rev. Lett. {\bf 99}, 173201 (2007).
\bibitem{Gau09} A. Gauguet, B. Canuel, T. L\'{e}veque, W. Chaibi, and A. Landragin, ¡°Characterization and limits of a cold-atom sagnac interferometer,¡± Phys. Rev. A {\bf 80}, 063604 (2009).
\bibitem{Sto11} J. Stockton, K. Takase, and M. Kasevich, ¡°Absolute geodetic rotation measurement using atom interferometry,¡± Phys. Rev. Lett. {\bf 107}, 133001 (2011).
\bibitem{Tac12} G. Tackmann, P. Berg, C. Schubert, S. Abend, M. Gilowski, W. Ertmer, and E. Rasel, ¡°Self-alignment of a compact large-area atomic sagnac interferometer,¡± New J. Phys. {\bf 14}, 015002 (2012).
\bibitem{Rak14} A. V. Rakholia, H. J. McGuinness, and G. W. Biedermann, ¡°Dual-axis high-data-rate atom interferometer via cold ensemble exchange,¡± Phys. Rev. Applied. {\bf 2}, 054012 (2014).
\bibitem{Davis08} J. Davis and F. Narducci, ¡°A proposal for a gradient magnetometer atom interferometer,¡± J. Mod. Opt. {\bf 55}, 3173-3185 (2008).
\bibitem{Zhou10} M.-K. Zhou, Z.-K. Hu, X.-C. Duan, B.-L. Sun, J.-B. Zhao, and J. Luo, ¡°Precisely mapping the magnetic field gradient in vacuum with an atom interferometer,¡± Phys. Rev. A {\bf 82}, 061602 (2010).
\bibitem{Hu11} Z.-K. Hu, X.-C. Duan, M.-K. Zhou, B.-L. Sun, J.-B. Zhao, M.-M. Huang, and J. Luo, ¡°Simultaneous differential measurement of a magnetic-field gradient by atom interferometry using double fountains,¡± Phys. Rev. A {\bf 84}, 013620 (2011).
\bibitem{Bar11} B. Barrett, I. Chan, and A. Kumarakrishnan, ¡°Atom-interferometric techniques for measuring uniform magnetic field gradients and gravitational acceleration,¡± Phys. Rev. A {\bf 84}, 063623 (2011).
\bibitem{Fix07} J. B. Fixler, G. Foster, J. McGuirk, and M. Kasevich, ¡°Atom interferometer measurement of the newtonian constant of gravity,¡± Science {\bf 315}, 74-77 (2007).
\bibitem{Lam08} G. Lamporesi, A. Bertoldi, L. Cacciapuoti, M. Prevedelli, and G. Tino, ¡°Determination of the newtonian gravitational constant using atom interferometry,¡± Phys. Rev. Lett. {\bf 100}, 050801 (2008).
\bibitem{Rosi14} G. Rosi, F. Sorrentino, L. Cacciapuoti, M. Prevedelli, and G. Tino, ¡°Precision measurement of the newtonian gravitational constant using cold atoms,¡± Nature {\bf 510}, 518-521 (2014).
\bibitem{Fray04} S. Fray, C. A. Diez, T. W. H\"{a}nsch, and M. Weitz, ¡°Atomic interferometer with amplitude gratings of light and its applications to atom based tests of the equivalence principle,¡± Phys. Rev. Lett. {\bf 93}, 240404 (2004).
\bibitem{Bon13} A. Bonnin, N. Zahzam, Y. Bidel, and A. Bresson, ¡°Simultaneous dualspecies matter-wave accelerometer,¡± Phys. Rev. A {\bf 88}, 043615 (2013).
\bibitem{Bon15} A. Bonnin, N. Zahzam, Y. Bidel, and A. Bresson, ¡°Characterization of a simultaneous dual-species atom interferometer for a quantum test of the weak equivalence principle,¡± Phys. Rev. A {\bf 92}, 023626 (2015).
\bibitem{Sch14} D. Schlippert, J. Hartwig, H. Albers, L. L. Richardson, C. Schubert, A. Roura, W. P. Schleich, W. Ertmer, and E. M. Rasel, ¡°Quantum test of the universality of free fall,¡± Phys. Rev. Lett. {\bf 112}, 203002 (2014).
\bibitem{Tar14} M. Tarallo, T. Mazzoni, N. Poli, D. Sutyrin, X. Zhang, and G. Tino, ¡°Test of einstein equivalence principle for 0-spin and half-integer-spin atoms: Search for spin-gravity coupling effects,¡± Phys. Rev. Lett. {\bf 113}, 023005 (2014).
\bibitem{Zhou15} L. Zhou, S. Long, B. Tang, X. Chen, F. Gao,W. Peng,W. Duan, J. Zhong, Z. Xiong, J. Wang et al., ¡°Test of equivalence principle at $10^{-8}$ level by a dual-species double-diffraction raman atom interferometer,¡± Phys. Rev. Lett. {\bf 115}, 013004 (2015).
\bibitem{Wil04} J. G. Williams, S. G. Turyshev, and D. H. Boggs, ¡°Progress in lunar laser ranging tests of relativistic gravity,¡± Phys. Rev. Lett. {\bf 93}, 261101 (2004).
\bibitem{Sch08} S. Schlamminger, K.-Y. Choi, T. Wagner, J. Gundlach, and E. Adelberger, ¡°Test of the equivalence principle using a rotating torsion balance,¡± Phys. Rev. Lett. {\bf 100}, 041101 (2008).
\bibitem{Col75} R. Colella, A. W. Overhauser, and S. A. Werner, ¡°Observation of gravitationally induced quantum interference,¡± Phys. Rev. Lett. {\bf 34}, 1472 (1975).
\bibitem{Lam98} C. L?mmerzahl, ¡°Quantum tests of the foundations of general relativity,¡± Class. Quantum. Grav. {\bf 15}, 13 (1998).
\bibitem{Dim08} S. Dimopoulos, P. W. Graham, J. M. Hogan, and M. A. Kasevich, ¡°General relativistic effects in atom interferometry,¡± Phys. Rev. D {\bf 78}, 042003 (2008).
\bibitem{Duan15} X.-C. Duan, X.-B. Deng, M.-K. Zhou, W.-J. Xu, F. Xiong, Y.-Y. Xu, C.-G. Shao, J. Luo, and Z.-K. Hu, ¡°Test of the universality of free fall with atoms in different spin orientations,¡± arXiv preprint arXiv:1602.06377 (2016).
\bibitem{Sug13} A. Sugarbaker, S. M. Dickerson, J. M. Hogan, D. M. Johnson, and M. A. Kasevich, ¡°Enhanced atom interferometer readout through the application of phase shear,¡± Phys. Rev. Lett. {\bf 111}, 113002 (2013).
\bibitem{Kuhn14} C. Kuhn, G. McDonald, K. Hardman, S. Bennetts, P. Everitt, P. Altin, J. Debs, J. Close, and N. Robins, ¡°A bose-condensed, simultaneous dualspecies mach¨Czehnder atom interferometer,¡± New J. Phys. {\bf 16}, 073035 (2014).
\bibitem{Agu10} D. Aguilera, H. Ahlers, B. Battelier, A. Bawamia, A. Bertoldi, R. Bondarescu, K. Bongs, P. Bouyer, C. Braxmaier, L. Cacciapuoti et al., ¡°STE-QUEST-test of the universality of free fall using cold atom interferometry,¡± Class. Quantum. Grav. {\bf 31}, 115010 (2014).
\bibitem{Alt15} B. Altschul, Q. G. Bailey, L. Blanchet, K. Bongs, P. Bouyer, L. Cacciapuoti, S. Capozziello, N. Gaaloul, D. Giulini, J. Hartwig et al., ¡°Quantum tests of the einstein equivalence principle with the ste¨Cquest space mission,¡± Adv. Space Res. {\bf 55}, 501-524 (2015).
\bibitem{Fos02} G. Foster, J. Fixler, J. McGuirk, and M. Kasevich, ¡°Method of phase extraction between coupled atom interferometers using ellipse-specific fitting,¡± Opt. Lett. {\bf 27}, 951-953 (2002).
\bibitem{Sto07} J. K. Stockton, X. Wu, and M. A. Kasevich, ¡°Bayesian estimation of differential interferometer phase,¡± Phys. Rev. A {\bf 76}, 033613 (2007).
\bibitem{Var09} G. Varoquaux, R. A. Nyman, R. Geiger, P. Cheinet, A. Landragin, and P. Bouyer, ¡°How to estimate the differential acceleration in a two-species atom interferometer to test the equivalence principle,¡± New J. Phys. {\bf 11}, 113010 (2009).
\bibitem{Chen14} X. Chen, J. Zhong, H. Song, L. Zhu, J. Wang, and M. Zhan, ¡°Proportional-scanning-phase method to suppress the vibrational noise in nonisotope dual-atom-interferometer-based weak-equivalence-principletest experiments,¡± Phys. Rev. A {\bf 90}, 023609 (2014).
\bibitem{Per15} F. P. Dos Santos, ¡°Differential phase extraction in an atom gradiometer,¡± Phys. Rev. A {\bf 91}, 063615 (2015).
\bibitem{Bar15} B. Barrett, L. Antoni-Micollier, L. Chichet, B. Battelier, P.-A. Gominet, A. Bertoldi, P. Bouyer, and A. Landragin, ¡°Correlative methods for dualspecies quantum tests of the weak equivalence principle,¡± New J. Phys. {\bf 17}, 085010 (2009).
\bibitem{Cla95} A. Clairon, P. Laurent, G. Santarelli, S. Ghezali, S. Lea, and M. Bahoura, ¡°A cesium fountain frequency standard: preliminary results,¡± IEEE Trans. Instrum. Meas. {\bf 44}, 128-131 (1995).
\bibitem{Che06} P. Cheinet, F. P. Dos Santos, T. Petelski, J. Le Gou\"{e}t, J. Kim, K. Therkildsen, A. Clairon, and A. Landragin, ¡°Compact laser system for atom interferometry,¡± Appl. Phys. B {\bf 84}, 643-646 (2006).
\bibitem{Mer09} S. Merlet, J. Le Gou\"{e}t, Q. Bodart, A. Clairon, A. Landragin, F. P. Dos Santos, and P. Rouchon, ¡°Operating an atom interferometer beyond its linear range,¡± Metrologia {\bf 46}, 87 (2009).
\bibitem{Zhou13} M.-K. Zhou, B. Pelle, A. Hilico, and F. P. dos Santos, ¡°Atomic multiwave interferometer in an optical lattice,¡± Phys. Rev. A {\bf 88}, 013604 (2013).
\bibitem{Li15} X. Li, C.-G. Shao, and Z.-K. Hu, ¡°Raman pulse duration effect in highprecision atom interferometry gravimeters,¡± J. Opt. Soc. Am. B {\bf 32}, 248-257 (2015).
\bibitem{Che08} P. Cheinet, B. Canuel, F. P. D. Santos, A. Gauguet, F. Yver-Leduc, and A. Landragin, ¡°Measurement of the sensitivity function in a timedomain atomic interferometer,¡± IEEE Trans. Instrum. Meas. {\bf 57}, 1141-1148 (2008).

\end{thebibliography}
\end{document}